\documentclass[12pt,aps,prd,showpacs,amsmath,amssymb]{revtex4}
\usepackage{slashed,amsmath,float}
\usepackage{mathrsfs}
\usepackage{amsfonts}
\usepackage{graphicx}
\usepackage{color}
\usepackage{ulem}
\usepackage{subfigure}
\begin{document}
\title{Electromagnetic form factors of $\Lambda_c$ in the Bethe-Salpeter equation approach }

\author{Liang-Liang Liu $^{a}$}
\email{corresponding  author. liu06_04@sxnu.edu.cn}
\author{Chao Wang $^{b}$}
\email{chaowang@nwpu.edu.cn}
\author{ Xin-Heng Guo $^{c}$}
\email{ corresponding  author. xhguo@bnu.edu.cn}
\affiliation{\footnotesize (a)~College of Physics and information engineering, Shanxi Normal University, Linfen 041004, People's Republic of China}
\affiliation {\footnotesize (b)~Center for Ecological and Environmental Sciences, Key Laboratory for Space Bioscience and Biotechnology, Northwestern Polytechnical University, Xi$^\prime$an, 710072, People's Republic of China}
\affiliation{\footnotesize (c)~College of Nuclear Science and Technology, Beijing Normal University, Beijing 100875, People's Republic of China}
\begin{abstract}
We study the electromagnetic form factors (EMFFs) of $\Lambda_c$ and the quark and diquark current contributes to the EMFFs of $\Lambda_c$ in the space-like (SL) region in the Bethe-Salpeter equation approach.
In this picture, the heavy baryon $\Lambda_c $ is regarded as composed of a heavy quark and a scalar diquark.
We find that for different values of parameters the quark and diquark current contribute to the EMFFs of $\Lambda_c$ is very different, but the total contribute to the EMFFs of $\Lambda_c$ is similarly.
The EMFFs of $\Lambda_c$ are similar to those of other baryons (proton, $\Xi^-$, $\Sigma^+$) with a peak at $\omega =1$ ($\omega=v^\prime \cdot v $ is the velocity transfer between the initial state (with velocity $v$) and the final state (with velocity $v^\prime$) of $\Lambda_c$).
\end{abstract}


\maketitle

\section{Introduction}

The quark-diquark model has been successful in describing nucleon properties \cite{Anselmino}.
A fully relativistic description of baryons can be accomplished in an approach in which baryons are considered as bound states of diquarks and quarks.
Ref.\cite{Gernot} has given a detailed overview of the quark-diquark model and EMFFS for the nucleon and $\Delta$ baryon.
In this reference the author gives the properties of diquark in different models.
Evidence for correlated diquark states in baryons was found in deep-inelastic lepton scattering \cite{Donnachie,Frederiksson,PMW} and in hyperon weak decays \cite{Stech}.
Attempts have been made to describe diquarks and baryons in non-local approximations to QCD \cite{Burden}. Diquark bound states were studied in Ref. \cite{Thorsson}.
The diquark EMFFs in a Nambu-Jona-Lasinio model were studied in \cite{Weiss}.
Spin-1 diquark contribution to the formation of tetraquarks in light mesons was studied in \cite{Hungchong}.
The properties of diquark in the rainbow-ladder framework was studied in  \cite{Gernot2}.

The nucleon EMFFs describe the spatial distributions of electric charge and current inside the nucleon and they are intimately related to nucleon internal structure.
They are not only  important observable parameters but also a essential key to understand the strong interaction \cite{Arrington, Perdrisat}.
In the past two decades, some theoretical investigations about EMFFs in both space-like (SL) and time-like (TL) regions \cite{JPBC0, JPBC1, Earle, ADGS,JXU,J-U} and  a lot of experimental results on EMFFs of baryons \cite{Bourgeois, Jlab, Walk, Arr, Bost, Lung, VoL, Horn, Tade, Cau, Kub, GDK, BESIII} and mesons \cite{Len, Col, Fra, J.R.Green} have appeared.
The SL region EMFFs of $\Lambda$ and $\Sigma$ were calculated in the framework of light-cone sum rule (LCSR) up to twist 6 \cite{YL-L,HMQ}. It was found that the $Q^2$-dependent magnetic form factor of $\Lambda$ approaches zero faster than the dipole formula with the increase of $Q^2$.

In previous work \cite{liang, Zhang-L, Guo-XH, Liu Y, Wu-HK, Weng-Mh}, we studied some properties of $\Lambda_b$ in the quark and diquark model.
In the present paper we will study the EMFFs of $\Lambda_c$ in the quark-diquark picture and calculate the contributions of quark and diquark currents to the EMFFs of $\Lambda_c$ in the SL region in the Bethe-Salpeter (BS) equation approach.
In our model, $\Lambda_c$ is regarded as a bound state of two particles: one is a heavy quark and the other is a $(ud)$ diquark.
This model has been successful in describing some baryons \cite{H.Meyer,A. De Ruijula,G. Karl,F. Close}.
In this picture, the BS equation for $\Lambda_b $ has been studied
extensively \cite{liang, Zhang-L, Guo-XH, Liu Y, Wu-HK, Weng-Mh}.
Similarly, $\Lambda_c$ can be described as $c(u d)_{0 0}$ (the first and second subscripts correspond to the spin and the isospin of the $(u d)$ diquark, respectively).
Then with the covariant instantaneous approximation and applying the kernel which includes the scalar confinement and the one-gluon-exchange terms, we will calculate the EMFFs of $\Lambda_c$.

The paper is organized as follows. In Section II, we will establish the BS equation for $\Lambda_c $ as a bound state of $c(u d)_{0 0}$.
In Section III we will derive the EMFFs for $\Lambda_c$ in the BS equation approach.
In Section IV the numerical results for the EMFFs of $\Lambda_c $ will be given.
Finally, the summary and discussion will be given in Section V.

\section{BS EQUATION FOR $\Lambda_c$}\label{sec2}

Generally, the BS wave function of $c(ud)_{0 0}$ system can be defined as the folowing \cite{liang, Guo-XH, Zhang-L, Liu Y, Wu-HK}:
\begin{eqnarray}\label{chi-x}
  \chi(x_1,x_2,P) &=& \langle0|T\psi(x_1) \varphi(x_2)|P\rangle,
\end{eqnarray}
where $\psi(x_1)$ and $\varphi(x_2)$ are the field operators of the $c$-quark and $(ud)_{00}$ diquark, respectively, $P=M v$ is the momentum of $\Lambda_c$.
We use $M,~m_c,\text{and}~m_D$ to represent the masses of the $\Lambda_c$, the $c$-quark and the $(ud)$ diquark, respectively, and $v$ to represent $\Lambda_c$'s velocity.
We define the BS wave function in momentum space:
\begin{eqnarray}\label{chi-f}
  \chi(x_1,x_2,P) &=& e^{i P X}\int \frac{d^4 p}{(2\pi)^4}e^{i p x} \chi_P(p),
\end{eqnarray}
where $X= \lambda_1 x_1+\lambda_2 x_2$ is the coordinate of center mass, $\lambda_1=\frac{m_c}{m_c+m_D} $, $\lambda_2=\frac{m_D}{m_c+m_D} $, and $x= x_1-x_2$. In momentum space, the BS equation for the $c(ud)_{00}$ system satisfies the homogeneous integral equation \cite{liang, chao, Liu Y, Guo-XH, Wu-HK, Weng-Mh, Zhang-L}
\begin{figure}[!ht]
\begin{center}
\includegraphics[width=9.5cm] {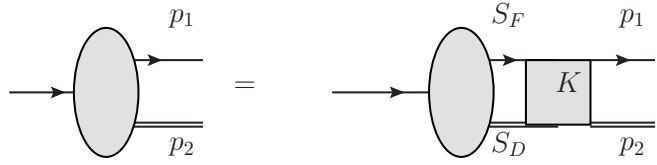}
\caption{The BS equation for $c(ud)_{00}$ system in momentum space (K is the interaction kernel)}\label{BSE}
\end{center}
\end{figure}

\begin{eqnarray}\label{chi-p}
\chi_P(p) &=& i S_F(p_1)\int \frac{d^4 q}{(2 \pi)^4} \left[ I\otimes I V_1(p,q)+\gamma_\mu \otimes \Gamma^\mu V_2(p,q)  \right]\chi_P(q)S_D(p_2),
\end{eqnarray}
where the quark momentum $p_1=\lambda_1 P+p$ and the diquark momentum $p_2=\lambda_2 P-p$, $S_F(p_1)$ and $S_D(p_2)$ are propagators of the quark and the scalar diquark, respectively,  $\Gamma^\mu=(p_2+q_2)^\mu \frac{\alpha_{seff}Q_0^2}{Q^2+Q^2_0}$ is introduced to describe the structure of the scalar diquark \cite{Guo-XH,M.Ansel, GS}, and $Q_0^2$ is a parameter in the form factor of the diquark which is related to the overlap integral of diquark wave functions. When $Q^2 \ll Q_0^2$ the form factor is frozen and when $Q^2 \gg Q_0^2$ the form factor can be determined perturbatively.
By analyzing the EMFFS of proton, it was found that $Q_0^2=3.2$ GeV$^2$ can lead to consistent results with the experimental data \cite{M.Ansel}.
It was found that the value of $Q_0^2$ is the order of $1$ GeV$^2$ in different model \cite{liang}.
$V_1$ and  $V_2$ are the scalar confinement and one-gluon-exchange terms.
It has been shown that in the quark-diquark model the $c(ud)_{00}$ system needs two scalar functions to describe the BS wave function \cite{liang, Liu Y}
\begin{eqnarray}
  \chi_P(p) &=& (f_1(p_t^2)+\slashed{p}_t f_2(p_t^2))u(v),
\end{eqnarray}
where $f_1, ~f_2$ are the Lorentz-scalar functions of $p_t^2$, $u(v)$ is the spinor of $\Lambda_c$, $p_t$ is the transverse projection of the relative momenta along the momentum $P$, $p_t^\mu = p^\mu-p_l v^\mu$ and $p_l=v \cdot p $.
According to the potential model, $V_1$ and $V_2$ have the following forms in the covariant instantaneous approximation ( $p_l=q_l$) \cite{Guo-XH,Zhang-L,Weng-Mh,Wei-Kw}:
\begin{eqnarray}\label{V1}
  \tilde{V}_1(p_t-q_t) &=& \frac{8 \pi \kappa}{[(p_t-q_t)^2+\mu^2]^2}-(2\pi)^2\delta^3(p_t-q_t)\int \frac{d^3k}{(2\pi)^3} \frac{8 \pi \kappa}{(k^2+\mu^2)^2},
\end{eqnarray}

\begin{eqnarray}\label{V2}
  \tilde{V_2} (p_t-q_t)&=&- \frac{16 \pi }{3}\frac{\alpha_{seff} Q^2_0}{(p_t-q_t)^2+\mu^2},
\end{eqnarray}
where $q_t$ is the transverse projection of the relative momenta along the momentum $P$ and defined as $ q_t^\mu = q^\mu -q_l v^\mu$, $q_l=v \cdot q$. The second term of $\tilde{V}_1$ is introduced to avoid infrared divergence at the point $ p_t=q_t$, $\mu$ is a small parameter to avoid the divergence in numerical calculations.
The parameters $\kappa$ and $\alpha_{seff}$ are related to scalar confinement and the one-gluon-exchange diagram, respectively.
For mesons the parameter of scalar confinement $\kappa^\prime$ is around $0.2$ GeV$^2$, but for baryons the dimension of the parameter $\kappa$ is three, the extra dimension in $\kappa$ should be caused by nonperburbative diagrams which include the frozen form factor at low momentum region.
Since $\Lambda_{QCD}$ is the only parameter which is related to confinement, we expect that $\kappa \sim \Lambda_{QCD} \kappa^\prime$, so the parameter $\kappa$ should be the order of $0.01$ GeV$^3$.
By analyzing the average kinetic energy of $\Lambda_b$ \cite{Wu-HK}, it was found the range of $\kappa$ is from $0.02$ to $0.08$GeV$^3$.
Therefore, in our numerical calculations we will take $\kappa$ to be in this range.

The quark and diquark propagators can be written as the following:
\begin{eqnarray}\label{SF}
  S_F(p_1)&=&i \slashed{v} \left[ \frac{\Lambda^+_c}{\lambda_1 M +p_l -\omega_c +i \epsilon}+\frac{\Lambda_c^-}{\lambda_1 M +p_l +\omega_c -i \epsilon}\right],
\end{eqnarray}
\begin{eqnarray}\label{SD}
  S_D(p_2) &=&\frac{i}{2 \omega_D} \left[\frac{1}{\lambda_2 M-p_l-\omega_D+i \epsilon}-\frac{1}{\lambda_2M-p_l+ \omega_D-i\epsilon}\right],
\end{eqnarray}
where $\omega_c = \sqrt{m_c^2-p_t^2}~\text{and}~\omega_D = \sqrt{m_D^2-p_t^2} $.
$\Lambda^\pm_c=1/2 \pm  \slashed{v}(\slashed{p}_t+m_c)/(2\omega_c)$ are the projection operators which satisfy the relations, $\Lambda_c^\pm \Lambda_c^\pm = \Lambda^\pm_c,~ \Lambda^\pm_c \Lambda^\mp_c = 0$.

Defining $\tilde{f}_{1(2)}=\int \frac{d p_l}{2 \pi}f_{1(2)}$, and using the covariant instantaneous approximation, $p_l=q_l$, we find that the scalar BS wave functions satisfy the coupled integral equation as follows

\begin{eqnarray}\label{eig1}
\tilde{f}_1(p_t) =&& \int \frac{d^3q_t}{(2\pi)^3} \bigg\{ \bigg[\frac{(\omega_c  +m_c ) (\tilde{V}_1+ 2 \omega_D \tilde{V}_2)-   p _t \cdot ( p _t+ q _t) \tilde{V}_2}{4 \omega_D \omega_c(-M + \omega_D+ \omega_c)} \nonumber\\
&& -\frac{(\omega_c -m_c)(\tilde{V}_1- 2\omega_D \tilde{V}_2)+   p _t\cdot( p _t+ q _t)  \tilde{V}_2}{4 \omega_D \omega_c(M + \omega_D+ \omega_c)} \bigg] \tilde{f}_1(q_t) \nonumber\\
&&+ \bigg[ \frac{-  (\omega_c+m_c) ( q _t + p _t)\cdot q_t\tilde{V}_2 +  p _t\cdot q_t(\tilde{V}_1- 2 \omega_D \tilde{V}_2)}{4 \omega_D \omega_c(-M + \omega_D+ \omega_c)} \nonumber\\
&&- \frac{(m_c- \omega_c )  ( q _t + p _t)\cdot q _t \tilde{V}_2 -   p _t\cdot q _t  (\tilde{V}_1+ 2\omega_D \tilde{V}_2)}{4 \omega_D \omega_c(M + \omega_D+ \omega_c)} \bigg]\tilde{f}_2(q_t) \bigg\},
\end{eqnarray}

\begin{eqnarray}\label{eig2}
\tilde{f}_2(p_t) = && \int \frac{d^3q_t}{(2\pi)^3} \bigg\{  \bigg[ \frac{(\tilde{V}_1+ 2 \omega_D \tilde{V}_2)-( -\omega_c+m_c) \frac{( p _t+ q _t) \cdot p _t }{ p^2_t }\tilde{V}_2}{4 \omega_D \omega_c(-M + \omega_D+ \omega_c)}    \nonumber\\
&&- \frac{- (\tilde{V}_1- 2\omega_D \tilde{V}_2)+(\omega_c  + m_c)\frac{( p _t+ q _t)\cdot p _t }{ p^2_t } \tilde{V}_2)}{4 \omega_D \omega_c(M + \omega_D+ \omega_c)}   \bigg]\tilde{f}_1(q_t) \nonumber \\
&& + \bigg[  \frac{(m_c -\omega_c)( \tilde{V}_1+ 2  \omega_D \tilde{V}_2）  ) \frac{ p_t \cdot q_t}{ p^2_t } - (  q^2_t+  p_t \cdot q_t) \tilde{V}_2}{4 \omega_D \omega_c(-M + \omega_D+ \omega_c)}    \nonumber\\
&&-\frac{ (m_c+\omega_c) (-\tilde{V}_1- 2 \omega_D \tilde{V}_2）) \frac{p_t \cdot q_t}{p^2_t} + (  q^2_t+  p_t \cdot q_t)\tilde{V}_2)}{4 \omega_D \omega_c(M + \omega_D+ \omega_c)} \bigg] \tilde{f}_2(q_t) \bigg\}.
\end{eqnarray}

Generally, the BS wave function can be normalized in the condition of the covariant instantaneous approximation \cite{Liu Y,Wei-Kw}:
\begin{eqnarray}\label{BSNOR}
  i \delta^{i_1 i_2}_{j_1 j_2} \int \frac{d^4 q d^4 p}{(2\pi)^8}\bar{\chi}_P(p,s)\left[\frac{\partial}{\partial P_0}I_p(p,q)^{i_1 i_2 j_2 j_1}\right]\chi_P(q,s^\prime)=\delta_{s s^\prime},
\end{eqnarray}
where $i_{1(2)}$ and $j_{1(2)}$ represent the color indices of the quark and the diquark, respectively, $s^{(\prime)}$ is the spin index of the baryon $\Lambda_c$, $I_p(p,q)^{i_1 i_2 j_2 j_1}$ is the inverse of the four-point propagator written as follows

\begin{eqnarray}\label{IPNOR}
  I_p(p,q)^{i_1 i_2 j_2 j_1} =\delta^{i_1 j_1}\delta^{i_2 j_2} (2 \pi)^4 \delta^4(p-q)S^{ -1 }_q(p_1)S^{ -1 }_D(p_2).
\end{eqnarray}

\section{SL electromagnetic form factors of $\Lambda_c$ }

In general, the SL EMFFs of $\Lambda_c$ can be defined by the matrix element of the electromagnetic current between the baryon states \cite{liang, J.R.Green, YL-L,HMQ}:
\begin{equation}\label{JJJ}
\langle \Lambda_c(P',s')|j_\mu(x=0)|\Lambda_c(P,s)\rangle=\bar u_{\Lambda_c}(P',s')\left[\gamma_\mu F_1(Q^2)+i\frac{\sigma_{\mu\nu} q^\nu}{2M}F_2(Q^2)\right]u_{\Lambda_c}(P,s),
\end{equation}
where $u_{\Lambda_c}(P,s)$ denotes the Dirac spinor of $\Lambda_c$ with momentum $P$ and spin $s$,
 $F_1(Q^2)$ and $F_2(Q^2)$ are Dirac and Pauli form factors, respectively, $M$ is the mass of $\Lambda_c$, $Q^2=-q^2=-(P-P')^2$ is the squared momentum transfer, and $j_\mu$ is the electromagnetic current relevant to the baryon.

In particular, similar to the nucleus the form factors $F_1$ and $F_2$ have the following values  when $Q^2 \rightarrow 0$, which corresponds to the exchange of low virtuality photon
\begin{eqnarray}\label{EMNOR}
  F_1(0) &=& 1,   \\
  F_2(0)&=&\kappa_{\Lambda_c},
\end{eqnarray}
where $\kappa_{\Lambda_c}=\mu_{\Lambda_c}-1$ ($\mu_{\Lambda_c} $ is the magnetic momentum of $\Lambda_c$).
Generally, considering perturbative QCD and helicity, $F_1(Q^2)$ and $F_2(Q^2)$ have the following behaviors at high $Q^2$ \cite{G.P, GPL, A.Efr, V.L.C, I.G.A, V.A.A, C.E.C, VCM,ST}

\begin{eqnarray}
F_1 \sim  \frac{1}{Q^4}, ~
F_2 \sim \frac{1}{Q^6}.\label{FF}
\end{eqnarray}

The Dirac and Pauli form factors are related to the magnetic and electric form factors $G_M(Q^2)$ and $G_E(Q^2)$
\begin{eqnarray}
G_M(Q^2)&=&F_1(Q^2)+F_2(Q^2), \label{gem1}\\
G_E(Q^2)&=&F_1(Q^2)-\frac{Q^2}{4M^2}F_2(Q^2).\label{gem2}
\end{eqnarray}

 At small $Q^2$, $G_E$ and $G_M$ can be thought of as Fourier transforms of the charge and magnetic current densities of the baryon. However, at large momentum transfer this view does not apply. Considering Eqs. (\ref{FF} - \ref{gem2}), at the large momentum transfer $|G_E|/|G_M|$ should be a stable value.

\begin{figure}[!ht]
\begin{center}
\includegraphics[width=9.5cm] {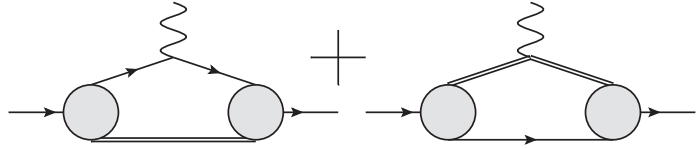}
\caption{The electromagnetic current is the sum of the quark current and the diquark current \cite{V.Keiner} }\label{diq}
\end{center}
\end{figure}

It is noted that Eq. (\ref{JJJ}) represents the microscopical description of the SL form factors of $\Lambda_c$ which include two contributions coming from the quark and the diquark, respectively, as is shown in Fig. \ref{diq}.
Therefore, in the quark-diquark model, the electromagnetic current $j_\mu$ coupling to $\Lambda_c$ is simply the sum of the quark and diquark currents.
So we have the relation  \cite{J.R.Green}:
\begin{eqnarray}\label{jj}
  j_\mu &=& j_\mu^{quark}+ j_\mu^{diquark},
\end{eqnarray}
where $j_\mu^{diquark}= \bar{D}\Gamma_\mu D $, $\Gamma_\mu$ is the vertex among the photon and the  diquarks which includes the scalar diquark form factor.
Considering the quark current contribution, we have
\begin{equation}\label{MJJ}
\langle \Lambda_c(v',s')|j_\mu^{quark}|\Lambda_c(v,s)\rangle=\bar u(v',s')[ g_{1q}(\omega)\gamma_\mu+g_{2q}(\omega)(v'+v)_\mu ]u(v,s),
\end{equation}
where $j_\mu^{quark}=\bar{c}\gamma_\mu c$, $v^{(')}=P^{(')}/M$ is the velocity of $\Lambda_c$, $\omega=v'\cdot v=\frac{Q^2}{2 M^2}+1$ is the velocity transfer, $g_{1q}$ and $g_{2q}$ are the functions of $\omega$ \cite{liang, Guo-XH,Liu Y,G-K}.
Similarly, considering the diquark current contribution we have
\begin{equation}\label{DJJ}
\langle \Lambda_c(v',s')|j_\mu^{diquark}|\Lambda_c(v,s)\rangle=\bar u(v',s')[ g_{1D}(\omega)\gamma_\mu+g_{2D}(\omega)(v'+v)_\mu ]u(v,s).
\end{equation}

 When $\omega=1$, we have the following relation \cite{Guo-XH}

\begin{eqnarray}\label{Nor}
  g_{1q}(1)+2 g_{2q}(1) &=& 1 +\mathcal{O}(1/M^2_{\Lambda_c}).
\end{eqnarray}

In the present work, we will use Eq. (\ref{Nor}) to normalize BS wave functions and neglect $1/M^2$ corrections \cite{G-K}. This relation has been proven to be a good approximation \cite{G-K} for a heavy baryon and proposed in \cite{M-B,M-V,H-M,BJM} for mesons. As shown in our previous works \cite{liang, Guo-XH}, we have
\begin{equation}\label{g1g2}
\langle \Lambda_c(v',s')|j_\mu |\Lambda_c(v,s)\rangle=\bar u(v',s')[ g_{1 }(Q^2)\gamma_\mu+g_{2 }(Q^2)(v'+v)_\mu ]u(v,s).
\end{equation}

Comparing Equations (\ref{g1g2}) and (\ref{JJJ}), we have
\begin{eqnarray}
g_{1}&=&F_1-\frac{F_2}{2},\label{g1q} \\
g_{2}&=&\frac{F_2}{4}.\label{g2q}
\end{eqnarray}

It can be shown that the matrix elements of the quark current and the diquark current can be written as the following:
\begin{eqnarray}\label{cq}
  \langle \Lambda_c(v',s')| j^{quark}_\mu(x=0)|\Lambda_c(v,s) \rangle &=&\int \frac{d^4 q}{(2 \pi)^4} \bar{\chi}(p') \gamma_\mu \chi (p) S_D^{-1}(p_2),\\
  \langle \Lambda_c(v',s')| j^{diquark}_\mu(x=0)|\Lambda_c(v,s) \rangle &=&\int \frac{d^4 q}{(2 \pi)^4} \bar{\chi}(p')\Gamma_\mu \chi (p) S_q^{-1}(p_1).
\end{eqnarray}

Considering the quark and diquark have same charge sign in the $c(ud)_{00}$ system we can calculate $g_1$ and $g_2$ as the following:
 \begin{eqnarray}
   g_1(\omega) &=& g_{1q}(\omega)+g_{1D}(\omega)\label{gg1} , \\
   g_2(\omega) &=& g_{2q}(\omega)+g_{2D}(\omega)\label{gg2} .
 \end{eqnarray}

Comparing Eqs. (\ref{MJJ}, \ref{DJJ}) and (\ref{gg1}, \ref{gg2}), we have:

\begin{eqnarray}\label{qdq1}
\bar{u}(v',s')[g_{1q}(\omega) \gamma_\mu + g_{2q}(\omega) (v' + v)_\mu]u(v,s)&=&\int \frac{d^4 q}{(2 \pi)^4} \bar{\chi}(p') \gamma_\mu \chi (p) S_D^{-1}(p_2),
\end{eqnarray}
\begin{eqnarray}\label{qdq2}
\bar{u}(v',s')[g_{1D}(\omega) \gamma_\mu + g_{2D}(\omega) (v' + v)_\mu]u(v,s)&=&\int \frac{d^4 q}{(2 \pi)^4} \bar{\chi}(p') \Gamma_\mu \chi (p) S_q^{-1}(p_1).
\end{eqnarray}

\section{Numerical analysis}

\subsection{Solution of the BS wave functions}
In order to solve Equations (\ref{eig1}, \ref{eig2}), we define the mass of $\Lambda_c$, $M=m_c+m_D+E$ where $E$ is the binding energy.
Taking $m_c=1.586$ GeV, $M=2.286$ GeV we have $ m_D+E=0.7$ GeV for $\Lambda_c$ \cite{Zhang-L}.
We choose the diquark mass $m_D$ to change from $0.83$ to $0.89$ GeV for $\Lambda_c$ so that the binding energy $E$ varies from $-0.2$ to $-0.1$ GeV.
Therefore, we choose the diquark mass $m_D$ to changes in the reasonable range from $0.83$ to $0.89$ GeV in our model.
The parameter $\kappa$ is taken to change from $0.02$ to $0.08$ GeV$^3$ \cite{Wu-HK}.
Hence, for each $m_D$, we can get a best value of $\alpha_{seff}$ corresponding to a value of $\kappa$ when solving Eqs. (\ref{eig1}, \ref{eig2}).

Solving the integral equations (\ref{eig1}, \ref{eig2}) we can get numerical solutions of the BS wave functions.
In Table \ref{TB1}, we give the values of $ \alpha_{seff}$ for $m_D=0.83,~0.86,~0.89$ GeV for different $\kappa$ when $Q_0^2=3.2$ GeV$^2$.
In Table \ref{TB2}, we give the values of $ \alpha_{seff}$ for $Q_0^2=1.0,~3.2,~10.0$ GeV$^2$ for different $\kappa$ when $m_D=0.86$ GeV.

\begin{table}[!htb]
\centering  
\begin{tabular}{l||c|c|c|c}  
\hline
  &$\alpha_{seff}(\kappa=0.02) $ &$\alpha_{seff}(\kappa=0.04 )$ &$\alpha_{seff}(\kappa=0.06 )$&$\alpha_{seff}(\kappa=0.08 )$ \\ \hline \hline
$m_D=0.83$ &0.78 &0.80 &0.84 &0.86 \\         
$m_D=0.86$ &0.80 &0.84 &0.86 &0.88 \\        
$m_D=0.89$ &0.84 &0.86 &0.88 &0.90\\ \hline \hline
\end{tabular}
\caption{When $Q^2_0=3.2$ GeV$^2$ the values of $\alpha_{seff}$  for $\Lambda_c$ with different $m_D$ (GeV) and $\kappa~(\text{GeV}^3)$.}\label{TB1}
\end{table}

\begin{table}[!htb]
\centering  
\begin{tabular}{l||c|c|c|c}  
\hline
  &$\alpha_{seff}(\kappa=0.02) $ &$\alpha_{seff}(\kappa=0.04 )$ &$\alpha_{seff}(\kappa=0.06 )$&$\alpha_{seff}(\kappa=0.08 )$ \\ \hline \hline
$Q^2_0=1.0 $ &0.82 &0.86 &0.88 &0.90 \\         
$Q^2_0=3.2 $ &0.76 &0.78 &0.80 &0.82 \\ 
$Q^2_0=10.0$ &0.72 &0.74 &0.76 &0.78\\ \hline \hline
\end{tabular}
\caption{When $m_D=0.86$ GeV the values of $\alpha_{seff}$  for $\Lambda_c$ with different $Q^2_0$ (GeV$^2$) and $\kappa~(\text{GeV}^3)$.}\label{TB2}
\end{table}

\begin{figure}[!htb]
\begin{center}
\includegraphics[width=9cm]{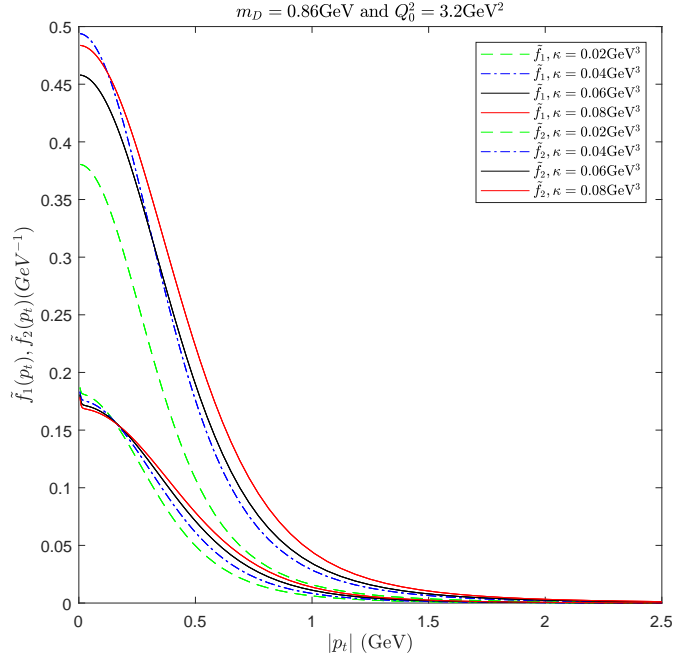}
\caption{(color online ) The BS wave functions for $\Lambda_c$ when $m_D=0.86$ GeV and $Q_0^2=3.2$ GeV$^2$.}\label{bsk}
\end{center}
\end{figure}

\begin{figure}[!htb]
\begin{center}
\includegraphics[width=9cm]{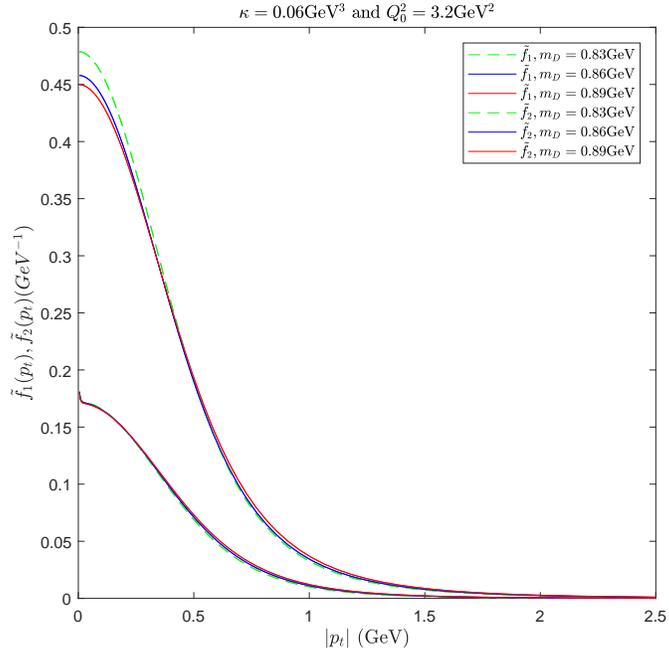}
\caption{(color online ) The BS wave functions for $\Lambda_c$ when $\kappa=0.06$ GeV$^3$ and $Q_0^2=3.2$ GeV$^2$.}\label{bsm}
\end{center}
\end{figure}

\begin{figure}[!htb]
\begin{center}
\includegraphics[width=9cm]{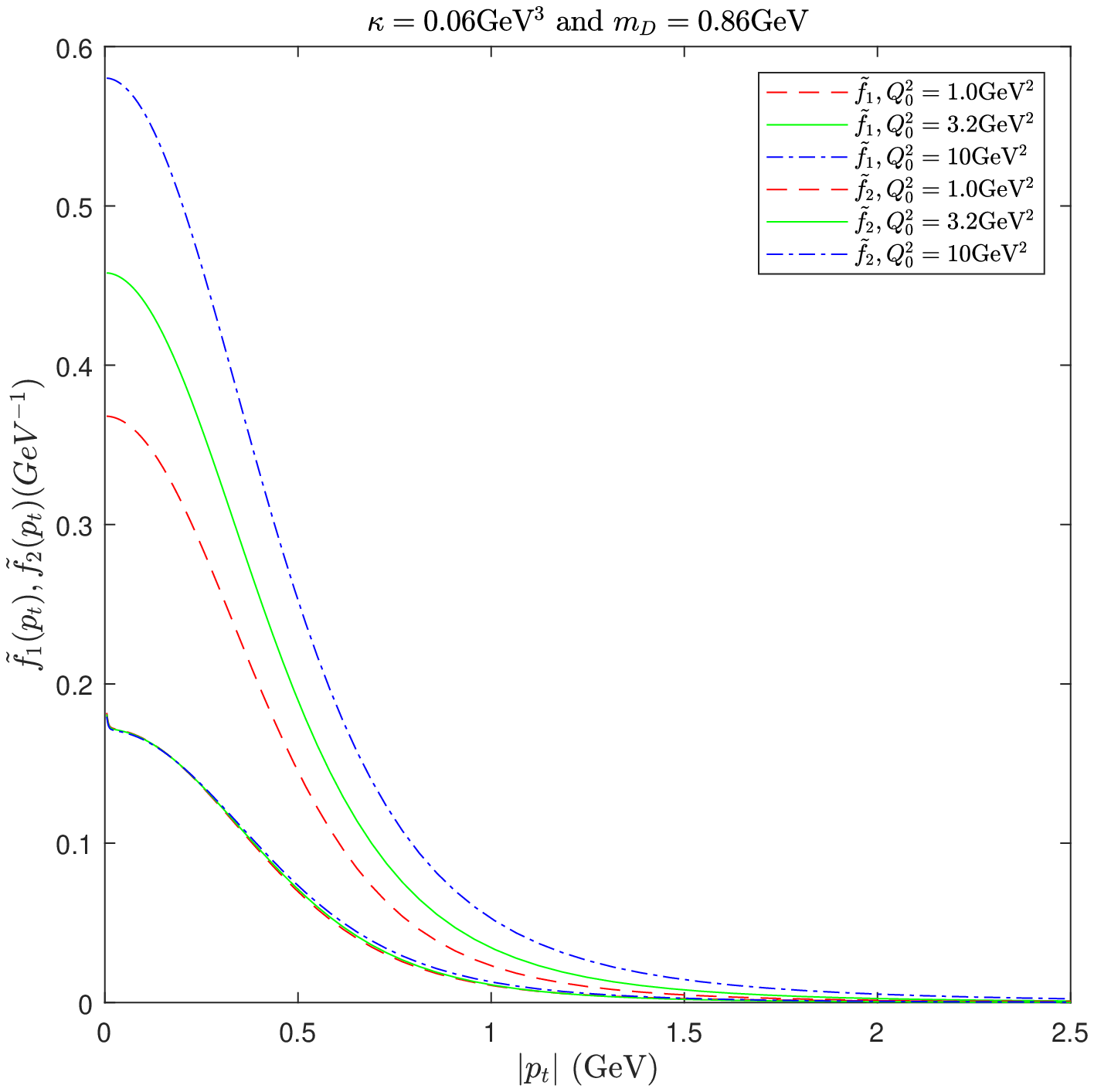}
\caption{(color online ) The BS wave functions for $\Lambda_c$ when $\kappa=0.06$ GeV$^3$ and $m_D=0.86$ GeV.}\label{bsq}
\end{center}
\end{figure}

In Figs. \ref{bsk} ,\ref{bsm}, \ref{bsq}, we plot $\tilde{f}_i~(i=1,2)$ depending on $|p_t|$.
 We can see from these figures that for different $\alpha_{seff}$ and $\kappa$, the shapes of BS wave functions are quite similar.
All the wave functions decrease to zero when $|p_t|$ is larger than about $2.0$ GeV due to the confinement interaction.
We find that the uncertainly of $m_D$ has a smaller impact on BS wave functions than that of $Q^2_0$ for the same value of $\kappa$.

\vskip 1cm

\subsection{Calculation of electromagnetic form factors of $\Lambda_c$}

In order to solve Eq. (\ref{qdq1}), we need the relations of $p$ and $p^\prime$.
We define $\theta$ to be the angle between $p_t$ and $v^\prime_t$ where $v^\prime_t=v^\prime-(v\cdot v^\prime)v$, then we have
\begin{eqnarray}
|v_t^\prime| &=& \sqrt{\omega^2-1} \label{vt},\\
 p_t \cdot v_t^\prime &=& - |p_t| |v_t^\prime | \cos\theta. \label{pvt}
\end{eqnarray}

Considering $p_2=p_2^\prime$, we obtain the following relations:
\begin{eqnarray}
p_t \cdot v_t^\prime &=&- |p_t| \sqrt{\omega^2-1} \cos\theta \label{pvtt}, \\
p^\prime_t \cdot v &=& p_l(1-\omega^2) + |p_t|\omega \sqrt{\omega^2-1} \cos\theta +  m_D( \omega-1 )^2 \label{ptv},\\
p_t \cdot p^\prime_t &=& (  p_l \omega - |p_t| \sqrt{\omega^2-1} \cos\theta -m_D \omega ) |p_t| \sqrt{\omega^2-1} \cos\theta  - |p_t| ^2  \label{pptt}.
\end{eqnarray}

Substituting Eqs. (\ref{SF}, \ref{SD}, \ref{vt} - \ref{pptt}) into  Eq. (\ref{qdq1}), integrating $p_l$ and using the relation $\tilde{f}^{\prime}_{1(2)}=\int \frac{d p^{\prime}_l}{2 \pi}f^{\prime}_{1(2)}$, $g_{1q},~g_{2q}$
can be expressed by $\tilde{f}^{(\prime)}_{(1,2)}$.
Similarly, for solving Eq. (\ref{qdq2}), we repeat the above process with $S_F^{-1}(p_1)$ being replaced by $S_D^{-1}(p_2)$ and replace the relation $p_2=p_2^\prime$ by $p_1=p_1^\prime$.

Substituting $g_{1q},~g_{2q},~g_{1D}$, and $g_{2D}$ into Eqs. (\ref{gem1}, \ref{gem2}) the EMFFs $G_E$ and $G_M$ can be written as
\begin{eqnarray}
  G_E &=& g_{1q}- 2 \omega (g_{2q}+g_{2D}) \label{ge1} ,\\
  G_M &=& g_{1q}+ 6 (g_{2q}+g_{2D}) \label{gm1}.
\end{eqnarray}

\vskip 0.5 in

\begin{figure}[!htb]
\begin{center}
\includegraphics[width=9cm]{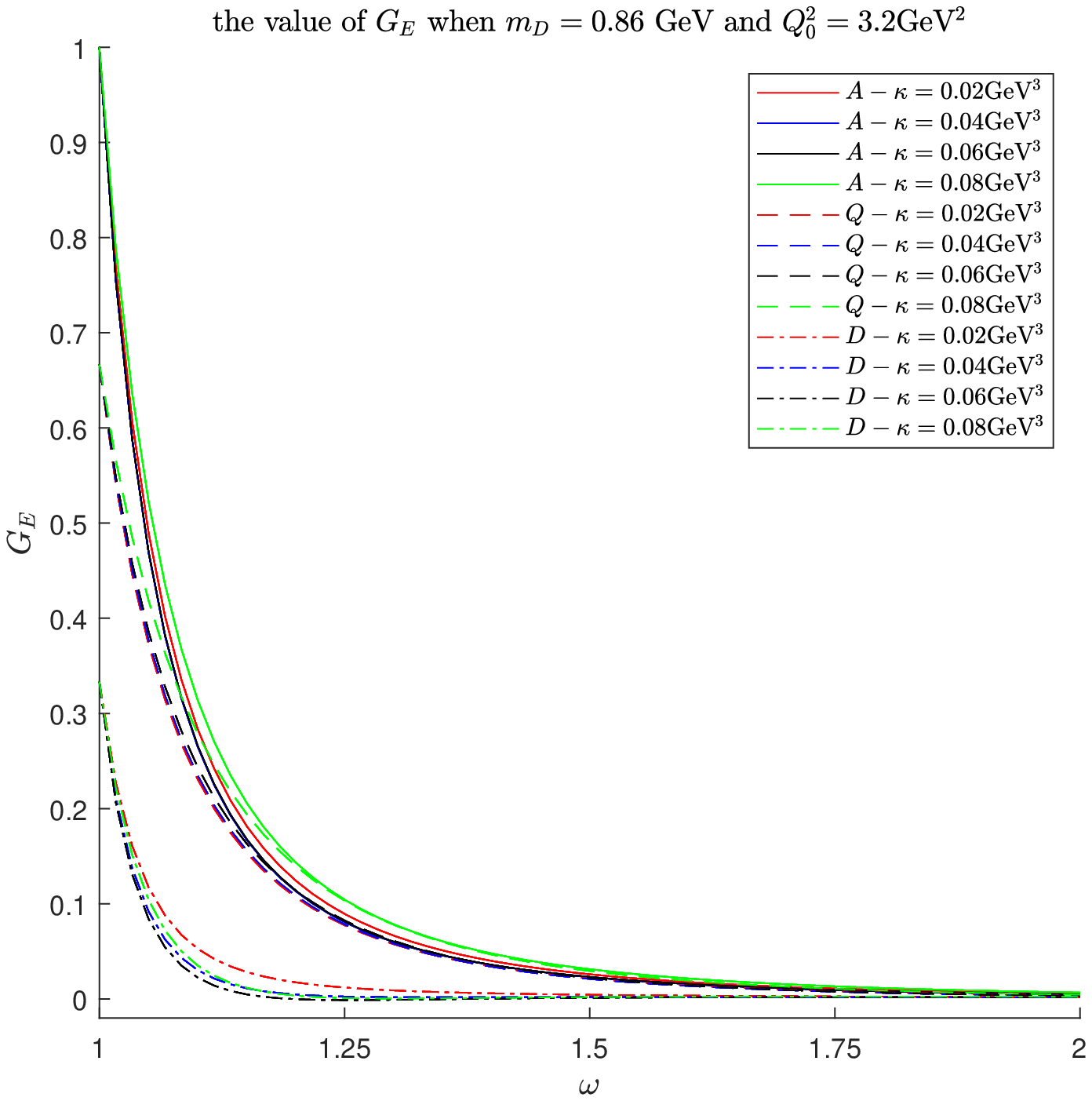}
\caption{(color online ) $\omega$-dependence of the electric form factor of $\Lambda_c$ for  $m_D=0.86$ GeV, $Q_0^2=3.2$ GeV$^2$ and different values of $\kappa$ ("Q" and "D" denote quark and diquark current contributions, respectively, "A" denotes the total contribution).}\label{GE75}
\end{center}
\end{figure}


\begin{figure}[!htb]
\begin{center}
\includegraphics[width=9cm]{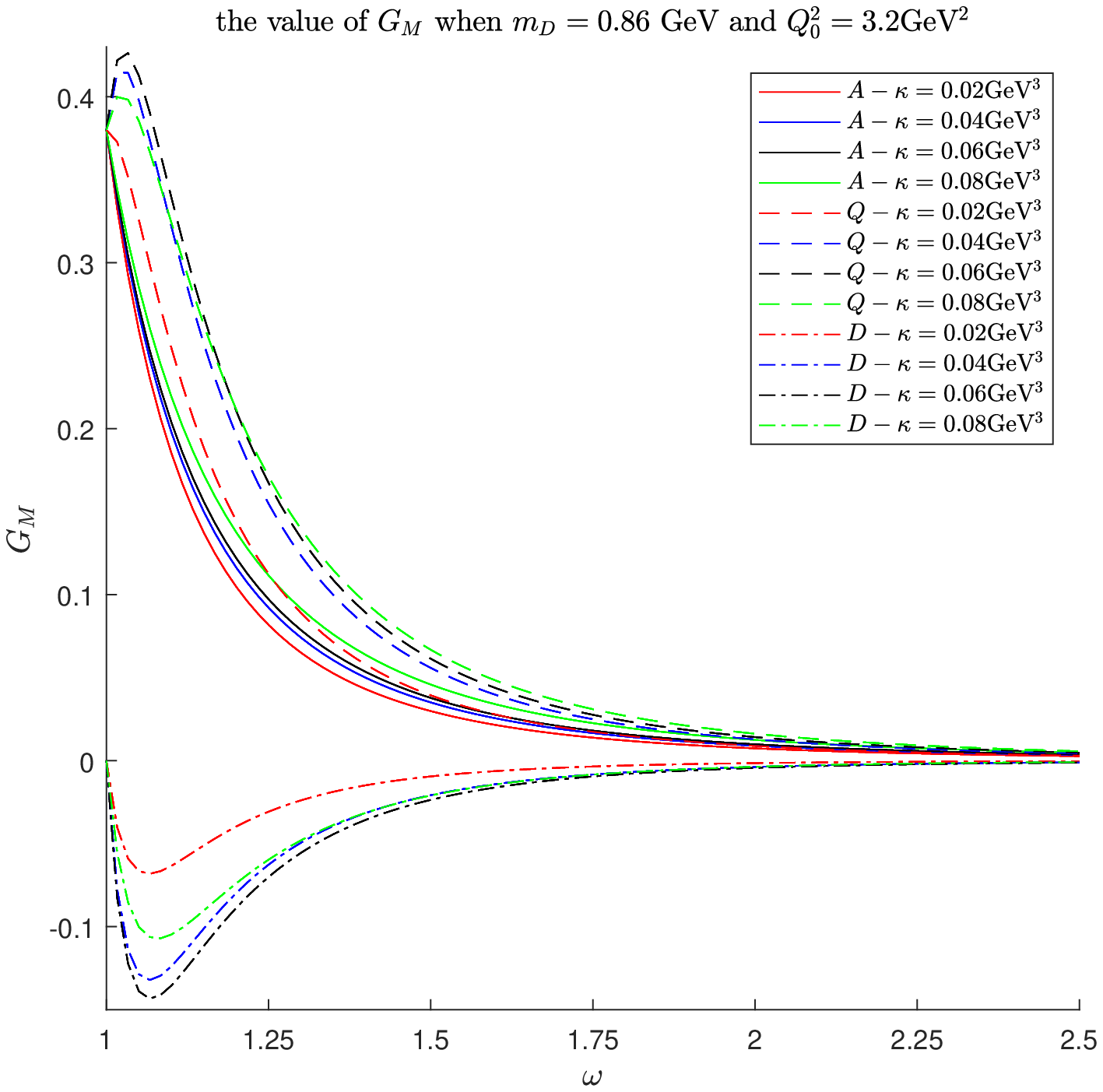}
\caption{(color online ) $\omega$-dependence of the magnetic form factor of $\Lambda_c$ for $m_D=0.86$ GeV, $Q_0^2=3.2$ GeV$^2$ and different values of $\kappa$ ("Q" and "D" denote quark and diquark current contributions, respectively, "A" denotes the total contribution).}\label{GM75}
\end{center}
\end{figure}%

\begin{figure}[!htb]
\begin{center}
\includegraphics[width=9cm]{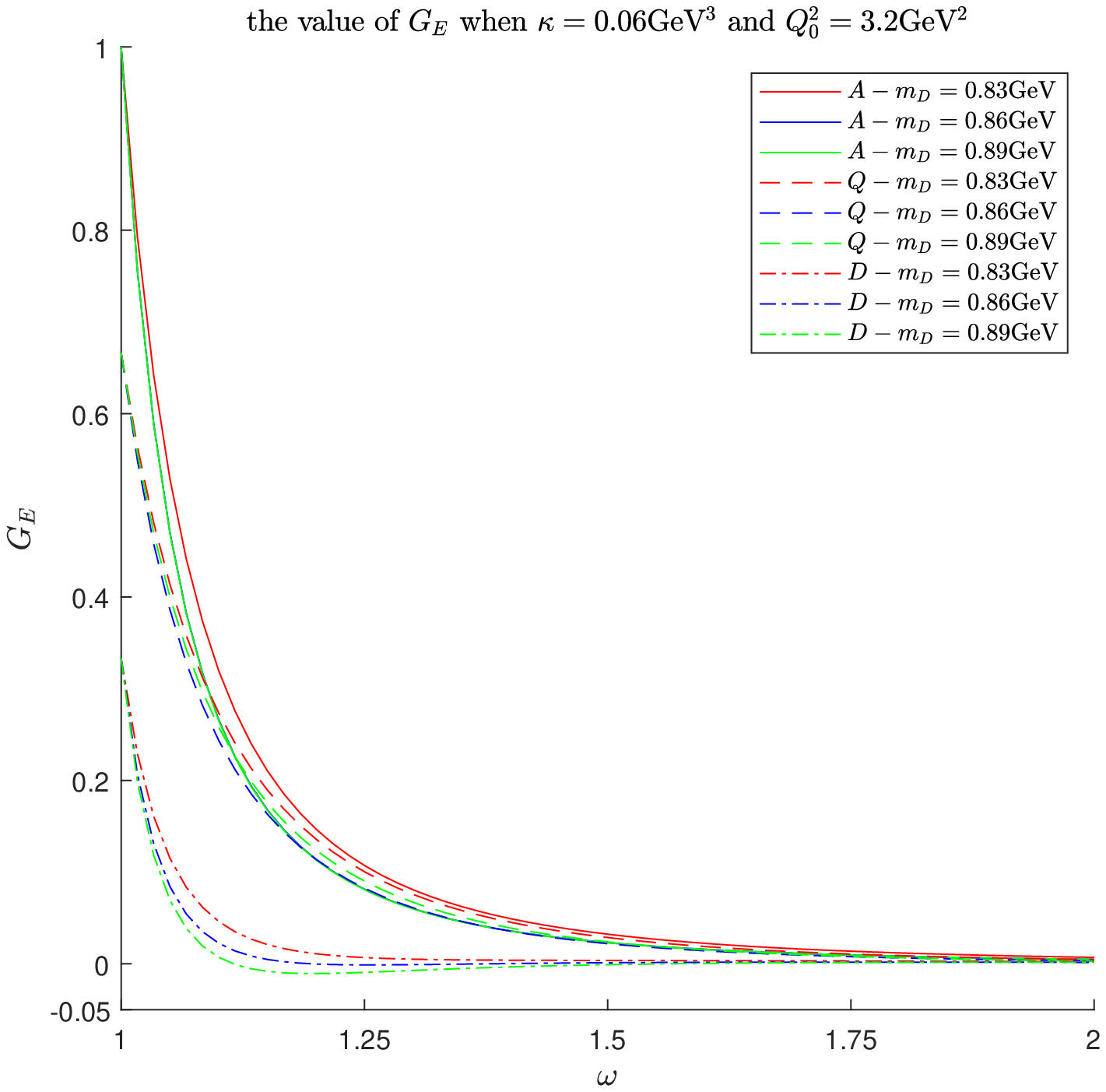}
\caption{(color online ) $\omega$-dependence of the electric form factor of $\Lambda_c$ for $\kappa=0.06$ GeV$^3$, $Q_0^2=3.2$ GeV$^2$  and different values of $m_D$ ("Q" and "D" denote quark and diquark current contributions, respectively, "A" denotes the total contribution). }\label{GEk6}
\end{center}
\end{figure}


\begin{figure}[!htb]
\begin{center}
\includegraphics[width=9cm]{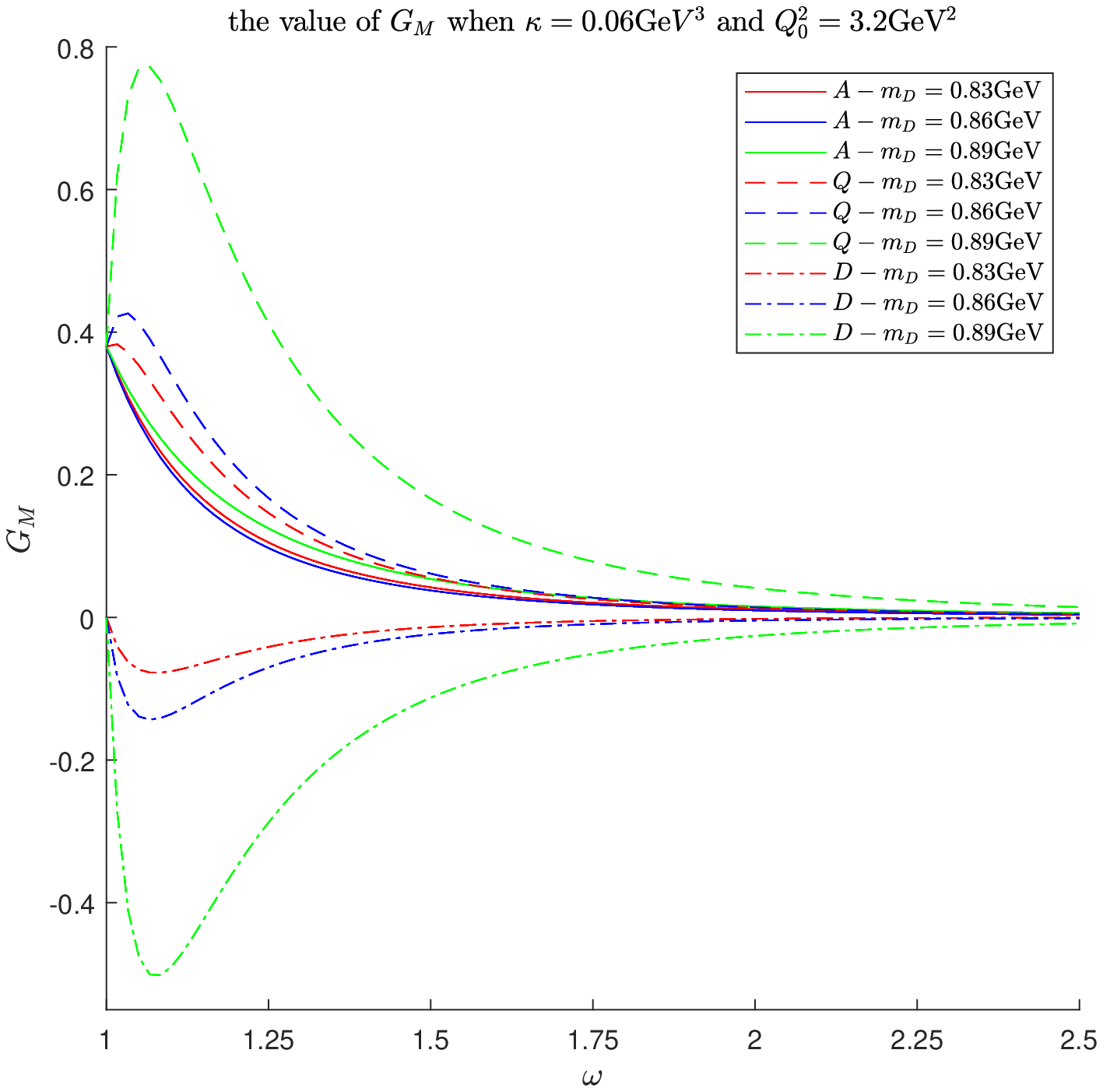}
\caption{(color online ) $\omega$-dependence of the magnetic form factor of $\Lambda_c$ for $\kappa=0.06$ GeV$^3$, $Q_0^2=3.2$ GeV$^2$  and different values of $m_D$ ("Q" and "D" denote quark and diquark current contributions, respectively, "A" denotes the total contribution).  }\label{GMk6}
\end{center}
\end{figure}


\begin{figure}[!htb]
\begin{center}
\includegraphics[width=9cm]{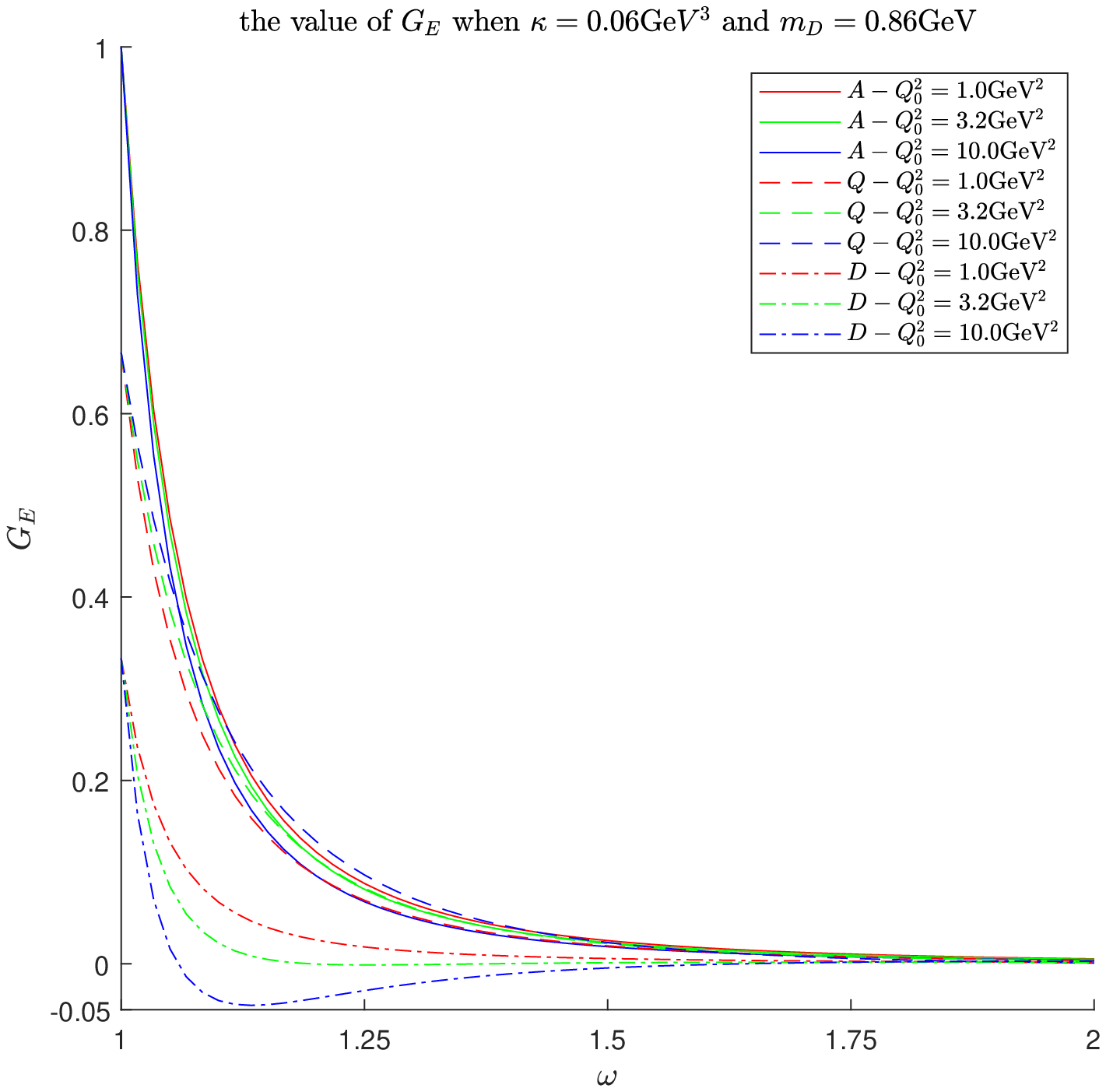}
\caption{(color online ) $\omega$-dependence of the electric form factor of $\Lambda_c$ for $\kappa=0.06$ GeV$^3$, $m_D=0.86$ GeV and different values of $Q_0^2$ ("Q" and "D" denote quark and diquark current contributions, respectively, "A" denotes the total contribution).  }\label{GEQ6}
\end{center}
\end{figure}


\begin{figure}[!htb]
\begin{center}
\includegraphics[width=9cm]{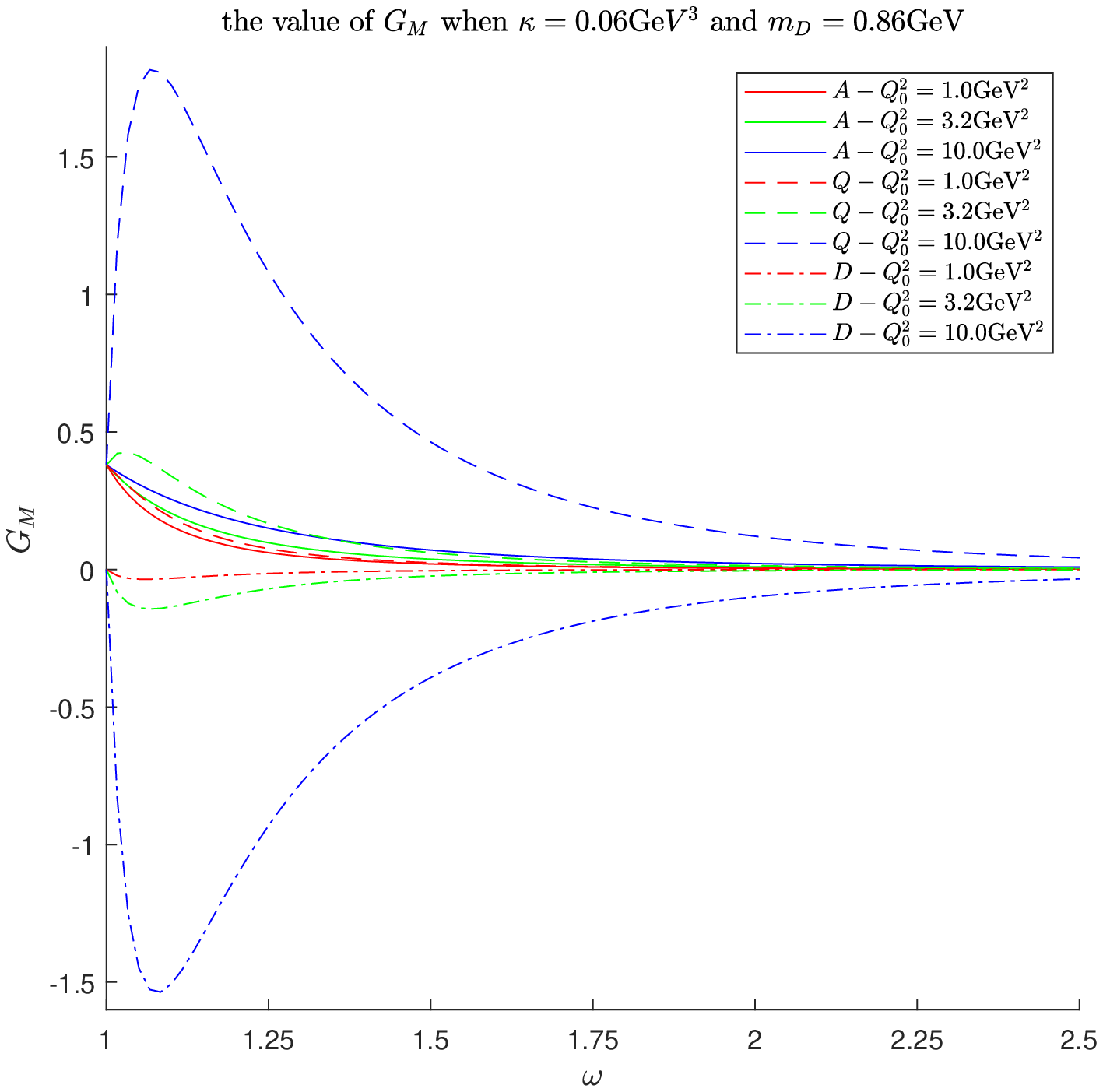}
\caption{(color online ) $\omega$-dependence of the magnetic form factor of $\Lambda_c$ for $\kappa=0.06$ GeV$^3$,$m_D=0.86$ GeV and different values of $Q_0^2$ ("Q" and "D" denote quark and diquark current contributions, respectively, "A" denotes the total contribution).  }\label{GMQ6}
\end{center}
\end{figure}

In Figs. \ref{GE75}-\ref{GMQ6}, we plot the $\omega$-dependence of $G_E$ and $G_M$ for different parameters.
From these figures, we find that for different $Q_0^2$, $m_D$ and $\kappa$, the shapes of $G_E$ and $G_M$ are similar.
In the range of $\omega$ from $1.0$ to $3.0$, the trends of $G_E$ and $G_M$ for $\Lambda_c$ are similar to those for the proton, $\Xi^-$, and $\Sigma^+$ \cite{YL-L, SAH}.

From these figures, we also find that $G_E$ decreases more rapidly than $G_M$ as $\omega$ increases.
For the electric form factors $G_E$, $\kappa$ causes the smallest uncertainly. However, for the magnetic form factors $G_M$, $m_D$ causes the smallest uncertainly.
This trend is different from $\Lambda_b$ \cite{liang}.
In the dipole model, $G_M(Q^2)=\frac{\mu}{(1+Q^2/m_0^2)^2}$, $\mu \propto 1/M$ (For $\Lambda_{(c)}$, $M$ is the mass of $s (c)$ quark) corresponds to the baryon  magnetic moment and for $\Lambda$, the parameter $m_0=\sqrt{0.89}$ GeV \cite{HMQ}.
There is no data for EMFFs of $\Lambda_c$ at present.
However, for $\Lambda$ and $\Lambda_c $ baryons the ratio of $|G_E|$ and $|G_M|$, $RM$, should be of order $M_s/M_c$.

\begin{equation}\label{guji}
RM=|\frac{G_{M_{\Lambda_c}} }{ G_{M_{\Lambda}}}| \propto \frac{M_s}{M_c}.
\end{equation}

For $\Lambda$ and $\Lambda_c$, the ratio $RM$ is about $0.3$ in the dipole model.
From Ref. \cite{YL-L} we know that the magnetic form factor of $\Lambda$ decreases faster than that in the dipole model.
Therefore, the ratio $RM$ should be the order of $0.1$.
In the range of $\omega$ from $1.0$ to $2.5$, our result for $|G_{M\Lambda_c}|$ varies from about $0.38$ to $0$.
In different models \cite{Ramalho2,  Ramalho, YL-L,SAH}, $|G_{M\Lambda}|$ varies from about $0.43 \sim 0.75$ to $0$.
Then we optain the ratio $RM$ to be about $0.26 \sim 0.47$.
For the magnetic moment of $\Lambda_c$ the traditional QCD sum rules \cite{zhu} gives the value $\mu_{\Lambda_c}=0.15 \pm 0.05 \mu_N $ ($\mu_N$ is the nucleon magnetic moment).
In the light cone QCD sum rules, Ref. \cite{aliev} gives $\mu_{\Lambda_c}=0.40 \pm 0.05 \mu_N $. In our model, obtain $\mu_{\Lambda_c} \approx 0.38\mu_N $.
These results agree roughly.

From Figs. \ref{GE75}-\ref{GMQ6} we find that the EMFFs of $\Lambda_c$ from quark and diquark current contributions are very different.
$\kappa$ leads to the smallest uncertainly and $Q^2_0$ leads to the largest impact.
From Figs. \ref{GE75} and \ref{GM75}, we find that for different $\kappa$ the EMFFs of $\Lambda_c$ primarily come from the quark contribution.
From Figs. \ref{GEQ6} and \ref{GMQ6}, we find that for different $Q^2_0$ the contributions of quark and diquark currents are very different. However, we find that the total contributions of quark and diquark currents to the EMFFs of $\Lambda_c$ do not change a lot comparing with Figs. \ref{GE75}-\ref{GMk6}.

\section{summary and discussion}

In the quark-diquark model, $\Lambda_c$ is regarded as a bound state of $c$-quark and scalar diquark.
In this picture, we established the BS equation for $\Lambda_c$.
Then we solved the BS equation numerically by applying the kernel which includes the scalar confinement and the one-gluon-exchange terms.
Then, we calculated the EMFFs of $\Lambda_c$ including both the $c$- quark and the $(ud)_{00}$ diquark current contributes.

Lastly, we compared our results with those of other baryons.
We found that the shapes of the EMFFs of $\Lambda_c$ are similar to those of other baryons \cite{YL-L, Ramalho, Ramalho2, SAH}.
For different values of $m_D$ and $\kappa$ the electric form factor of $\Lambda_c$ changes in the range $1.0 \sim 0 $ as $\omega$ changes form $1.0$ to $2.0$ and the magnetic form factor of $\Lambda_c$ changes in the range   $0.4 \sim 0 $ as $\omega$ changes form about $1.0$ to $2.5$.
For different parameters, especial for $Q^2_0$, we found that the contributions of quark and diaquark currents are very different, but the total contributions of quark and diquark currents do not change a lot.

Depending on the parameters $m_D,  ~ \kappa~\text{and}~Q_0^2$ in our model, our results vary in some ranges.
We studied the uncertainties for $G_E$ and $G_M$ that can be caused by $\kappa$, $m_D$ and $Q_0^2$ and found that these uncertainties are less than $27\%$ due to $\kappa$, $20\%$ due to $m_D$ and $40\%$ due to $Q_0^2$.
Our results need to be tested in future experimental measurements.
In the future, our model can be used to study other baryons such as the proton, the neutron, $\Lambda$ and excited states of $\Lambda_{(b,c)}$.

\acknowledgments
This work was supported by National Natural Science Foundation of China under contract numbers 11775024, 11575023 and the Fundamental Research Funds for the Central Universities of China (Project No.~31020170QD052).


\end{document}